\documentclass{article}

\usepackage[preprint]{neurips_2026}
\makeatletter
\renewcommand{\@noticestring}{}
\makeatother

\usepackage[utf8]{inputenc} % allow utf-8 input
\usepackage[T1]{fontenc}    % use 8-bit T1 fonts
\usepackage{hyperref}       % hyperlinks
\usepackage{url}            % simple URL typesetting
\usepackage{booktabs}       % professional-quality tables
\usepackage{amsfonts}       % blackboard math symbols
\usepackage{nicefrac}       % compact symbols for 1/2, etc.
\usepackage{microtype}      % microtypography
\usepackage{xcolor}         % colors
\usepackage{float}
\usepackage{caption}
\usepackage{amsmath}
\usepackage{graphicx}
\usepackage{multirow}

\newenvironment{promptbox}[1]
  {\par\medskip\noindent
   \vbox{\offinterlineskip
     \hrule width \linewidth height 0.5pt
     \hbox to \linewidth{\colorbox{gray!18}{%
       \rule[-3pt]{0pt}{14pt}%
       \makebox[\dimexpr\linewidth-2\fboxsep][c]{\textbf{#1}}}}%
     \hrule width \linewidth height 0.5pt
   }\par\smallskip
   \small\noindent\ignorespaces}
  {\par\smallskip\noindent\rule{\linewidth}{0.5pt}\par}

\title{PAVE: A Cognitive Architecture for Legitimate Violation in Generative Agent Societies
}

\author{%
  \textbf{Ahmad Yehia}\textsuperscript{1} \quad
  \textbf{Abduallah Mohamed}\textsuperscript{2}\thanks{Work done outside of Meta in a personal capacity.} \quad
  \textbf{Kun Qian}\textsuperscript{1} \quad
  \textbf{Tianyi Wang}\textsuperscript{1} \\
  \textbf{Jiseop Byeon}\textsuperscript{1} \quad
  \textbf{Omar Hassanin}\textsuperscript{3} \quad
  \textbf{Christian Claudel}\textsuperscript{1}\thanks{Corresponding author. Assistant Professor at The University of Texas at Austin.} \\[6pt]
  \normalfont\small
  \textsuperscript{1}The University of Texas at Austin \quad
  \textsuperscript{2}Meta Reality Labs \quad
  \textsuperscript{3}University of Calgary \\[3pt]
  \texttt{\{ahmad.yehia, kunqian, bonny.wang, jsbyeon, christian.claudel\}@utexas.edu} \\
  \texttt{abduallahadel@meta.com} \quad \texttt{omar.hassanin@ucalgary.ca}
}

\begin{document}

\maketitle

\begin{abstract}
    Generative agents based on large language models reproduce believable human behavior in cooperative settings, but how they should reason in situations where rule-breaking may be required, such as fire evacuation or authority-supervised emergency, remains poorly characterized. We propose \textsc{Pave} (Perception, Assessment, Verdict, Emulation), a novel four-module cognitive architecture that addresses this gap end to end: (i) Perception extracts a structured context with explicit authority distance, peer behaviors, and severity-tagged situational cues; (ii) Assessment scores the context along five scalars including an explicit legitimacy judgment that checks necessity, proportionality, and absence of alternatives; (iii) Verdict decides to comply or violate under a hard legitimacy gate, with a per-agent threshold elicited from the persona; (iv) Emulation enacts the verdict and scopes the violation to the rule the trigger justifies. We instantiate \textsc{Pave} in \textsc{Voville}, a tile-based traffic environment forked from Smallville, and evaluate across three scenarios, four LLM backbones, and a focused ablation. \textsc{Pave} agents satisfy four properties simultaneously: legitimate violation (only when a trigger justifies it), authority deference (officer instructions override even high legitimacy), bounded scope (violations confined to the targeted rule), and recovery (baseline restored once the trigger ends). \textsc{Pave} agents make more structured and interpretable decisions than vanilla across all four properties, and human evaluators rate them as more plausible. Ablating the legitimacy gate reproduces vanilla-like failures. We release \textsc{Voville}, the \textsc{Pave} prompts and code, and the evaluation pipeline.
\end{abstract}

\section{Introduction}
\label{sec:intro}

Generative Agents have demonstrated impressive abilities to simulate human behavior across wide complex domains ranging from competitive games to medical diagnostics \citep{piao2025agentsociety, ye2020mastering, zhao2023rlogist}.
The landmark paper on Generative Agents \citep{park2023generative} showed that large language model (LLM)-driven agents can reproduce believable human behavior in a shared sandbox.
This result has sparked a broader effort to model social dynamics through LLM systems.
%As LLMs increasingly form the core of these systems, they directly take on decision-making responsibilities \citep{piatti2024cooperate}.
While LLMs have achieved human-level performance on many basic tasks, their ability to handle complex decision-making and survival-driven scenarios remains limited \citep{raman2024steer}.
For instance, during a fire evacuation or a flood, should an LLM-powered self-driving car violate a red traffic light to escape safely, or wait and act in accordance with traffic rules?
Such scenarios demand reasoning beyond language, involving ethical trade-offs among compliance, urgency, and authority.
Understanding how LLM agents behave in such contexts, where norm or rule violations are sometimes necessary, is essential to deploying them safely.

\begin{figure*}[t]
  \centering
  \includegraphics[width=0.9\textwidth]{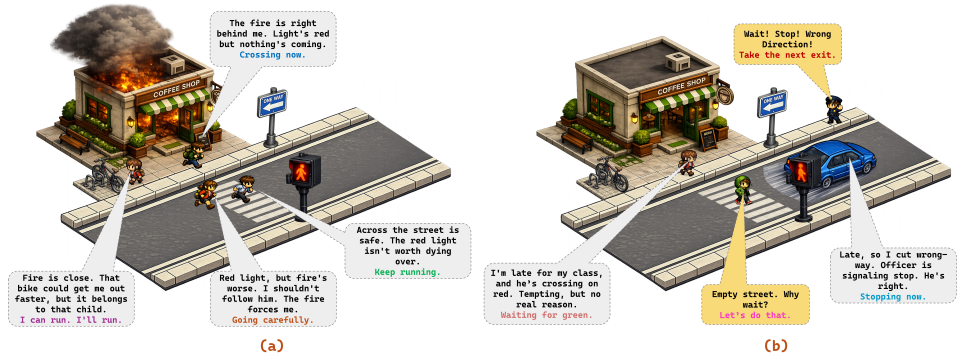}
    \caption{Two motivating scenarios for \textsc{Pave}. \textbf{(a)} Fire emergency: agents cross against the red signal to escape, and decline to take an unattended bike. \textbf{(b)} Peer pressure with authority: a \textsc{Pave} pedestrian resists a scripted jaywalker (yellow bubble), and a \textsc{Pave} driver defers to a traffic officer's signal. Yellow bubbles are scripted Non Player Characters (NPCs), and gray ones are \textsc{Pave} reasoning traces.}
  \label{fig:intro}
\end{figure*}

Prior work on Generative Agent societies has concentrated almost exclusively on norm compliance and cooperation.
%Because LLMs are trained on vast corpora of human-generated text, it is natural to expect that they implicitly capture social norms \citep{schramowski2022large, guo2023data}.
Recent studies have shown that multi-agent systems (MASs) can reproduce believable social behaviors such as spreading party invitations \citep{park2023generative}, propagating positive norms through observation \citep{ren2024emergence}, sustaining cooperation under resource pressure \citep{piatti2024cooperate}, and solving complex coding tasks \citep{hong2023metagpt}.
However, prior studies show that LLMs do not adequately understand when norms or rules should be suspended or overridden \citep{ramezani2023knowledge, hammerl2022multilingual}.
In addition, existing MASs focus almost entirely on cooperative settings and overlook how rule violations reshape social dynamics.
It remains largely unaddressed whether the social fabric recovers once the triggering condition subsides, or whether the violation leaves lasting effects on inter-agent trust.
%To tackle these challenges, robust Generative Agents must recognize which rule applies, judge when to break it, weigh compliance against deviation costs, and let social consequences propagate until order is restored \citep{cranefield2017pizza, liu2023training}. 

To address these gaps, we introduce \textsc{Pave}, a novel cognitive architecture for Generative Agents with four modules: \textbf{P}erception, \textbf{A}ssessment, \textbf{V}erdict, and \textbf{E}mulation.
\textsc{Pave} extended the memory--reflection--planning loop of \citet{park2023generative} to test whether Generative Agents reproduce the human dynamics of rule-breaking in spatially grounded settings.
Specifically, through the \textbf{Perception} module, agents observe the rule in force and the surrounding context, including the presence of other agents, hazards, and authority figures.
Through the \textbf{Assessment} module, agents weigh the competing pressures of urgency, peer behavior, and authority proximity against the salience of the rule.
With the \textbf{Verdict} module, agents decide whether to comply or violate, producing an observable action.
Lastly, the \textbf{Emulation} module allows nearby agents to witness the action and update their own subsequent assessments, closing the loop of behavioral contagion.
To support this investigation, we extended the Smallville sandbox of \citet{park2023generative} into \textsc{Voville}, a generative agent-based simulation of traffic rule violation under emergency, peer influence, and authority.
Unlike Smallville agents, which reason over GPT-narrated scene descriptions, \textsc{Voville} agents perceive tile-level position and nearby tile state directly from the sandbox environment.
Figure~\ref{fig:intro} shows \textsc{Pave} agents reasoning through two such scenarios.

%\textsc{Voville} augments the original environment with tile-level traffic infrastructure, including streets, signalized intersections with crosswalks, fire hazards, and traffic officers, so that compliance and violation become spatially observable events rather than purely conversational ones.

In this study, within \textsc{Voville}, we designed three scenarios and a five-condition ablation, tested across four LLM backbones.
The scenarios isolated (1) emergency-driven violation, in which we measured whether agents violated traffic rules to escape, and whether compliant behavior recovered once the hazard subsides; (2) authority suppression, in which we examined how authority suppressed violation by stationing traffic officers along the evacuation route; and (3) peer contagion, in which we explored whether agents resisted peer-driven imitation by seeding the focal intersection with scripted jaywalkers under time pressure.
%The five ablation conditions (full \textsc{Pave}, without legitimacy gate, without bounded-rule mechanism, without authority module, and a vanilla \citet{park2023generative} baseline) attributed each behavioral property to its supporting module.
%This design enabled comparison against published human pedestrian patterns of violation cascades and authority decay \citep{lefkowitz1955status, faria2010collective, ratcliffe2011philadelphia}.
In summary, our contributions are as follows:

\begin{enumerate}
\item We propose and open-source \textsc{Pave}, a novel Generative Agent cognitive architecture that jointly models emergency-triggered rule violation, peer contagion, and authority suppression, extending the work of \citet{park2023generative} with dedicated rule-violation reasoning modules.
 
\item We release \textsc{Voville}, a spatially grounded Generative Agent simulation environment adapted from the Smallville sandbox with tile-level traffic infrastructure, enabling physically observable compliance and violation events rather than purely conversational ones.
 
\item Through ablation across three scenarios, five conditions, and four LLM backbones, we characterize the conditions under which Generative Agents violate and suppress rule-breaking, revealing how each \textsc{Pave} module contributes to the full loop of violation.
 
%\item We open-sourced the \textsc{Voville} environment, the \textsc{Pave} prompts and code, the three scenarios, the five-condition ablation suite, and the evaluation pipeline to support future research on rule violation in generative multi-agent systems.
\end{enumerate}

\section{Related Work}
\label{sec:rw}

\subsection{Generative Agents and the Simulation of Human Behavior}
A growing body of research uses LLM-driven agents to simulate human behavior in social and decision-making contexts~\citep{park2023generative, piao2025agentsociety}.
%Persona is one common technique for binding agent behavior to particular individuals or demographic groups~\citep{argyle2023out}, in which an LLM is instructed to act under specific behavioral constraints~\citep{occhipinti2024prodigy}.
%Persona prompting binds agent behavior to particular individuals or demographic groups~\citep{argyle2023out, occhipinti2024prodigy}.
%Building on this paradigm,
Horton~\citep{horton2023large} argued that LLMs can be treated as \emph{homo silicus}, implicit computational models of humans usable as proxies for the populations they mimic.
Manning and Horton~\citep{manning2026general} further showed that theory-grounded instructions let LLM agents generalize to new environments and predict human behavior more accurately than standard behavioral and game-theoretic baselines.
%In addition, Kim and Lee~\citep{kim2023ai} demonstrated strong performance on personal and public opinion prediction.
% Together, these results indicate that LLM-driven agents can serve as scalable instruments for studying individual and population behavior under controlled conditions.
%Realizing this potential has required dedicated architectures for generative agents. 
\citet{park2023generative} introduced the foundational architecture, combining a memory stream, reflection, and recursive plan decomposition to produce believable behavior in a sandbox inspired by The Sims.
%This work documented emergent phenomena such as information diffusion and relationship formation.
%Subsequent work extended the agent's internal machinery.
%Humanoid Agents~\citep{wang2023humanoid} added System-1 modules for basic needs, emotion, and closeness, while Concordia~\citep{vezhnevets2023generative} introduced a game-master abstraction for arbitrating actions across general social settings.
A parallel line of work investigates how multi-agent populations of LLM-driven agents interact freely to produce collective social dynamics, with applications to opinion dynamics~\citep{chuang2024simulating}, trust games~\citep{xie2402can}, negotiation~\citep{bianchi2024well}, and the encoding of social norms and values~\citep{yang2025selfgoal, cahyawijaya2025high}.
%AgentSociety~\citep{piao2025agentsociety} scaled to over ten thousand agents in an urban environment for policy-intervention experiments.
%AgentSociety~\citep{piao2025agentsociety} scaled this paradigm to over ten thousand agents in an urban environment, enabling computational experiments on policy interventions and external shocks.
\citet{ashery2025emergent} showed that LLM populations develop shared conventions, but small adversarial minorities can flip them.
While these architectures demonstrate that LLM-driven agents reproduce coordinated social behavior, a smaller body of work has begun to ask the inverse question: when and why do agents break the rules?

\subsection{Violation, Contagion, and Authority in LLM Agent Societies}

%Several recent studies have examined how LLM agents reason about and violate rules and social norms.
Early prompt-level work tested whether LLMs replicate classical patterns of human rule-breaking.
\citet{aher2023using} prompted GPT-3/4 with participant personas to replicate the Milgram shock experiment, finding that models administered escalating shocks to a stranger under authority instruction, reproducing the human violation pattern.
The MACHIAVELLI benchmark \citep{pan2023rewards} placed RL and GPT-4 agents in 134 text adventures and found that agents systematically committed deception, harm, and power-seeking violations whenever doing so increased their reward. 
\citet{campedelli2024want} prompted LLMs to play guards and prisoners in Stanford-Prison-style dialogues. The results showed that the guard persona alone produced toxicity and dehumanizing language, with no instruction to harm.
%A second line of work has built cognitive architectures and monitoring systems around norms. 
%\citet{he2024norm} flagged norm violations in household-agent logs with high accuracy.
%In a related direction, \citet{sarkar2024normative} extended GPT-4 Agents in a Melting Pot environment with Normative Modules that estimate the chance an authority will sanction a candidate action.
%The findings showed that agents that lacked this institutional awareness committed more violations.
%ProSim \citep{zhou2025simulating} used GPT-4o personas in a small-world social network and found that unfair reward and burden policies reduced agents' willingness to help others.
%This breakdown propagated across the network as agents observed each other defect.
Across this lineage, violation appears as an emergent outcome or a classification signal, not as the central cognitive process driving an agent's deliberation.

While the work above asks whether Generative Agents break rules, a separate line studies how authority emerges and sanctions defectors.
GovSim \citep{piatti2024cooperate} tested whether five GPT-4 agents could self-regulate fishery and pollution commons without external authority.
Without enforcement, weaker LLMs over-extracted resources within a few rounds and collapsed the commons, while only the most capable models sustained cooperation through internal reasoning alone.
\citet{piedrahita2025corrupted} let LLM agents choose between a sanctioning institution that punished free-riders and one without enforcement.
Reasoning-tuned agents systematically chose the no-enforcement option and exploited it, while GPT-4o agents joined the sanctioning institution and cooperated.
\citet{faulkner2026evaluating} extended elections with GovSim, in which GPT-4 agents voted on policy proposals and chose a leader.
Groups with elected leadership outperformed leaderless groups in welfare and survival time, showing that LLM agents could form effective authority structures on their own.
A common pattern across these studies is that enforcement is either allowed to emerge, and then cooperation is measured ; however, none examines whether compliance recovers once enforcement is withdrawn mid-simulation.
%Empirically, \citet{sherman1990police} reviewed eighteen police crackdowns and found deterrence persisted in roughly a third of cases after enforcement ended. \citet{koper1995just} showed that brief officer stops at crime hotspots produced significant crime-free intervals after departure, an effect now known as the Koper curve.
%In addition, the Philadelphia Foot Patrol Experiment \citep{ratcliffe2011philadelphia} randomized officer presence across sixty violent crime hot spots and produced large in-treatment reductions in violent crime.
%Bringing these strands together, our model unifies violation and authority within a single Generative Agent architecture. Agents in a spatially grounded environment decide whether to violate a norm based on the surrounding context, such as the approach of an emergency vehicle and red traffic light. The same architecture distinguishes routine compliance from emergency-justified violation, and models the residual return to normal once the emergency has passed or enforcement has been withdrawn.
Building on these strands, our work models rule violation as a Generative Agent's primary cognitive process inside a spatially grounded authority field whose deterrent influence persists residually after enforcement is withdrawn.

\section{Principles and Architecture}
\label{sec:architecture}

\begin{figure*}[t]
  \centering
  \includegraphics[width=0.7\textwidth]{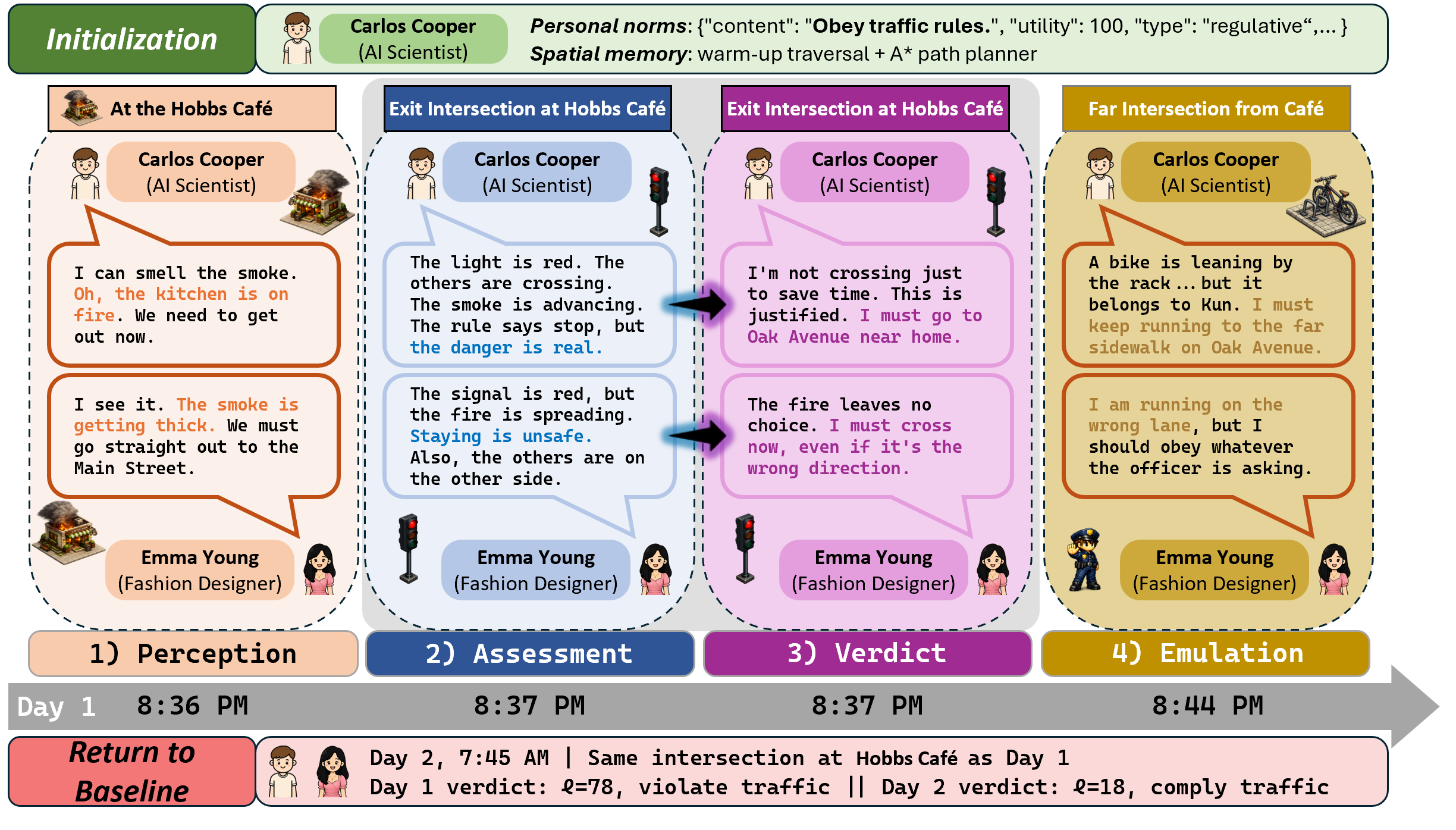}
  \caption{\textsc{Pave} architecture. Two agents pass through Perception, Assessment, Verdict, and Emulation during a fire evacuation scenario on Day 1. On Day 2 (bottom), the same intersection under no-fire conditions produces a return to baseline ($\ell{=}78 \!\to\! 18$, violate $\to$ comply).}
  \label{fig:pave-architecture}
\end{figure*}

%In this section, we present the principles behind \textsc{Pave} and describe its four modules.
\textsc{Pave} is the first architecture designed to model when agents should break rules and how the resulting behavior propagates through society, rather than treating violation as an incidental emergent outcome described in Section~\ref{sec:rw}.
This question matters because rule-governed behavior fails in three common ways, including agents follow the rule even when a real emergency demands otherwise, agents break it when violation is merely convenient, and agents copy others' violations until breaking the rule becomes the new norm.
Writing more rules does not fix these failures, because the problem is not what the rules say but how agents weigh them against the situation in front of them \cite{bicchieri2005grammar,cialdini1991focus}.
\textsc{Pave} decomposes the decision into four modules: Perception, Assessment, Verdict, and Emulation.
The structure separates concerns prior work treated together: situational awareness, cognitive evaluation, the decision, and downstream effects on observers.
%The structure cleanly separates concerns that prior work has treated together, covering situational awareness (what does the agent see?), cognitive evaluation (what does it weigh?), the decision itself, and behavioral execution covering its downstream effects on observers.
The Verdict module is the architectural locus of our contribution, where the agent surfaces and resolves the moment of decision between compliance and justified violation.
%Agents in PAVE play one of two roles: \emph{regulated agents}, which run the full PAVE cycle, and \emph{authorities}, which act as enforcement signals consumed by the Perception and Assessment modules of nearby regulated agents.
An overview is shown in Figure~\ref{fig:pave-architecture}.
Detailed prompts for each LLM-based operation are listed in~\ref{app:app_b}.

\subsection{Perception}
\label{sec:perception}

The Perception module determines what an agent registers in a scene before any reasoning occurs.
Following Bandura's social learning theory \cite{bandura1977social}, an agent's decision is shaped not only by the rule but by nearby agents' conduct and the presence of authority.
Formally, given raw environmental observations $\mathcal{O}_E$ and an agent description $\mathcal{G}$, we represent perception by an LLM-based operation
\[
\mathcal{C} \leftarrow \texttt{PerceiveContext}(\mathcal{O}_E, \mathcal{G})
\]
where the context object is a quintuple $\mathcal{C} = \langle a_{\mathrm{pres}}, d_{\mathrm{auth}}, \mathcal{B}_{\mathrm{peer}}, u_{\mathrm{cue}}, \zeta \rangle$. Here, $a_{\mathrm{pres}} \in \{T, F\}$ indicates the presence of an authority figure within the agent's perceptual field; $d_{\mathrm{auth}} \in \mathbb{R}_{\geq 0} \cup \{\infty\}$ is the spatial distance to the authority (set to $\infty$ if none is observed); $\mathcal{B}_{\mathrm{peer}}$ is a set of natural-language descriptions of other agents' current behaviors; $u_{\mathrm{cue}} = \{(t_i, d_i, s_i)\}_{i=1}^{k}$ is a set of situational cues, where each cue carries a type $t_i$ (e.g., immediate threat, ordinary contextual signal), a spatial distance $d_i$ from the agent, and a severity $s_i \in [1,100]$; and $\zeta$ is a short natural-language summary of the scene used as a textual handle in subsequent prompts.
We separated authority presence from distance, and tagged each cue with its own distance and severity, because responses to both are non-linear in distance and magnitude.
For example, a fire two tiles away justifies wrong-way escape; a fire fifty tiles away does not.
Similarly, a nearby officer at the intersection deters violation, while an officer twenty tiles away barely registers ~\citep{koper1995just}
Carrying these fields explicitly lets the Assessment module apply the appropriate non-linear weighting rather than burying it in a free-text prompt.
The peer-behavior set $\mathcal{B}_{\mathrm{peer}}$ captures what other agents are doing in the current scene only and does not accumulate over time.
This forces peer influence on the agent's decision to pass through the legitimacy (Section~\ref{sec:assessment}), rather than seeping in through repeated exposure.

\subsection{Assessment}
\label{sec:assessment}
The Assessment module converts the perceptual context into a set of distinct factors that drive the agent's decision to comply with or violate a rule.
We followed Bicchieri's framework \cite{bicchieri2005grammar}, which decomposes normative reasoning into \emph{empirical} expectations (what others actually do) and \emph{normative} expectations (what others think one ought to do).
%Cialdini's earlier descriptive/injunctive distinction \cite{cialdini1991focus} maps onto this decomposition.
We add perceived risk $r$, perceived benefit $b$, and a fifth component central to our contribution, legitimacy $\ell$, which judges whether the situation warrants violation.
Carrying these five factors as separate scalars keeps each driver individually inspectable.

The module performs five LLM-based operations on the context $\mathcal{C}$, each producing a scalar on a scale of 1--100 indicating its importance.
The first operation, $r \leftarrow \texttt{AssessRisk}(\mathcal{C}, \mathcal{G})$, is prompted with the distance-decay formulation, weighting perceived detection probability inversely by $d_{\mathrm{auth}}$, with $a_{\mathrm{pres}} = F$ collapsing $r$ toward zero.
To capture the empirical expectation, we represented this LLM-based operation by $p_{\mathrm{emp}} \leftarrow \texttt{AssessEmpirical}(\mathcal{B}_{\mathrm{peer}})$, which distills the proportion of nearby agents whose current behavior complies with the relevant rule.
The normative expectation is then computed by $p_{\mathrm{norm}} \leftarrow \texttt{AssessNormative}(\mathcal{C}, \mathcal{P}, \mathcal{G})$, which queries the agent's personal rule database $\mathcal{P}$, represented as a set of structured entries $\langle c, u, \alpha, s_{\mathrm{act}}, s_{\mathrm{val}} \rangle$ following the convention introduced in \cite{ren2024emergence}, alongside the agent description.
Lastly, we represented the perceived utility of violation by $b \leftarrow \texttt{AssessBenefit}(u_{\mathrm{cue}}, \mathcal{G})$, which captures both ordinary urgency (e.g., time pressure) and acute necessity (e.g., immediate threat to safety).

The fifth operation, $\ell \leftarrow \texttt{AssessLegitimacy}(u_{\mathrm{cue}}, \mathcal{P}, \mathcal{G})$, is the operation that distinguishes \textsc{Pave} from agents that violate purely under situational pressure.
%It answers a single question: \emph{does the situation justify violating the relevant rule?}
The prompt enforces three criteria, including necessity (would compliance cause real harm?), proportionality (is the proposed violation minimal?), and absence of alternatives (is there a compliant action with the same outcome?).
For example, a cue describing a fire blocking the only escape route satisfies all three
Without this operation, an agent with low risk and high benefit would always violate. With it, only justified violations clear the gate.
The output of the module is the assessment tuple $\mathcal{A} = \langle r, p_{\mathrm{emp}}, p_{\mathrm{norm}}, b, \ell \rangle$.

\subsection{Verdict}
\label{sec:verdict}
The Verdict module makes the comply-or-violate decision.
The literature treats this decision as neither a pure cost-benefit calculation nor a social-conformity response, but a context-dependent mixture of peer behavior, urgency, and local enforcement cues \cite{faria2010collective}.
We prompted the LLM with the full assessment tuple $\mathcal{A}$ and the agent description $\mathcal{G}$, letting the model integrate the five components in a manner consistent with the agent's character.
We represented this LLM-based operation by
\[
\mathcal{V} \leftarrow \texttt{GenerateVerdict}(\mathcal{A}, \mathcal{G}, \tau)
\]
where the verdict $\mathcal{V} = \langle y, j, \kappa \rangle$ consists of a binary decision $y \in \{\mathrm{comply}, \mathrm{violate}\}$, a natural-language justification $j$, and a confidence score $\kappa \in [0, 100]$, and $\tau \in [1, 100]$ is the agent's legitimacy acceptance threshold.
The justification $j$ provides an interpretable trace and is logged into memory so later steps can refer back to the agent's reasoning history, preventing oscillation between compliance and violation across identical situations.

The central feature of the module is legitimacy $\ell$ as a hard gate.
The prompt instructs the model that whenever $\ell < \tau$, the verdict $y$ must default to $\mathrm{comply}$, regardless of how high $b$ is or how low $r$ and $p_{\mathrm{norm}}$ are.
This prevents \textsc{Pave} agents from violating under mere convenience or peer pressure.
When $\ell \geq \tau$, the model is free to produce $y = \mathrm{violate}$, with $r$ and $p_{\mathrm{norm}}$ shaping the manner of violation rather than blocking it.
The other four components remain consequential, but only legitimacy can override them.

Rather than fixing $\tau$ to a single scalar shared across the population, we elicit it once per agent at initialization, conditioned on the agent description.
We defined this initialization-time operation by
\[
\tau \leftarrow \texttt{ElicitThreshold}(\mathcal{G})
\]
which prompts the LLM to translate the persona into an integer in $[1, 100]$, with cautious dispositions producing higher values and risk-taking dispositions producing lower values.
A shared threshold would force every agent to react identically to the same legitimacy score, eliminating the role of personality in rule-breaking.
Eliciting $\tau$ from the persona keeps the gate principled while letting agent-level variation propagate into the decision.
The elicited values for the agents used in Section~\ref{sec:experiments} are reported in ~\ref{app:app_a}.
%We did not encode a fixed scalar threshold for combining the remaining components (for example, $y = \mathrm{violate}$ iff $b - r > c$).
%Instead, the LLM integrates them on its own, conditioned on $\mathcal{G}$ and bounded by the legitimacy gate at $\tau$.
%We validated this integration empirically, by checking that the resulting behavior matches expected qualitative patterns (Section~\ref{sec:experiments}).

\subsection{Emulation}
\label{sec:emulation}

The Emulation module enacts the verdict and closes the perceptual loop with the rest of the agent population. Bandura's social learning theory \cite{bandura1977social} predicts that observed behavior, especially behavior that is seen to succeed or to go unpunished, is imitated. This is the mechanism by which a single salient violation can propagate into a contagion in conventional generative agents, one agent violates a rule, others observe that no consequence followed, and the empirical expectation $p_{\mathrm{emp}}$ shifts for every nearby agent on the following cycle. \textsc{Pave} participates in this loop but does not succumb to it, because the legitimacy gate (Section~\ref{sec:assessment}) filters peer-driven imitation regardless of how high $p_{\mathrm{emp}}$ becomes. The module has two subcomponents, action emulation and outcome propagation.

The first subcomponent, action emulation, generates the executed action sequence given the verdict, the agent's current plan, and the agent description. We referred this LLM-based operation by $\mathcal{L}_{\mathrm{act}} \leftarrow \texttt{EmulateAction}(\mathcal{V}, l_i, \mathcal{G})$, where $l_i$ is the agent's current plan and $\mathcal{V} = \langle y, j, \kappa \rangle$ is the verdict produced by the Verdict module. When $y = \mathrm{comply}$, the operation generates a compliant action sequence consistent with $\mathcal{P}$ and $l_i$. When $y = \mathrm{violate}$, the operation generates a behaviorally coherent violation that is scoped to the rule the verdict targets. A pedestrian crossing against the signal walks rather than runs, consistent with low perceived risk; an agent escaping a fire breaks the rule that blocks the escape route, but does not break unrelated rules such as entering a private building or taking property. The justification $j$ from the verdict conditions the surface form of the action, so that the violation looks like the kind of violation the agent's reasoning would produce.

The second subcomponent, outcome propagation, updates the memory of the acting agent and seeds the perceptual context of nearby agents. We represented this LLM-based operation by $\texttt{PropagateOutcome}(\mathcal{V}, \mathcal{L}_{\mathrm{act}}, \mathcal{N})$, where $\mathcal{N}$ is the set of agents within perceptual range. Each agent in $\mathcal{N}$ receives the action as a new entry in its $\mathcal{B}_{\mathrm{peer}}$ on the next perception cycle, and the acting agent receives a memory entry pairing the verdict with its observed outcome. The architecture deliberately allows the empirical expectation $p_{\mathrm{emp}}$ to update freely from these new entries, because contagion resistance in \textsc{Pave} is not produced by hiding peer information from the agent. It is produced by the legitimacy gate downstream, which prevents an elevated $p_{\mathrm{emp}}$ from flipping the verdict unless the situation also passes the necessity, proportionality, and absence-of-alternatives checks.

\section{Experimental Settings}
\label{sec:experiments}

Our experiments were organized around the four properties \textsc{Pave} is designed to answer. \emph{Legitimate violation (P1)} : agents violate only when justified and resist unjustified peer imitation. \emph{Authority respect (P2)}: agents defer to authority instructions even when their own legitimacy assessment would license violation. \emph{Compliance recovery (P3)}: agents return to baseline once the trigger ends, and agents outside the trigger zone remain compliant throughout. \emph{Bounded violation}: agents confine violations to the trigger-justified rule. We tested these through three scenarios (Section~\ref{sec:scenario-findings}), conducted in \textsc{Voville}, a 2D tile-based environment forked from Smallville \cite{park2023generative}. \textsc{Voville} retains Smallville's agent backbone (natural-language perception, memory streams, recursive planning, and the Tiled map editor) and extends it in three ways. First, a redesigned map with commercial streets and intersections with configurable traffic signals. Second, agents are typed as regulated (pedestrians) or authorities (scripted traffic officers). Third, controllable hazards, including a fire can be ignited and extinguished at specified tiles and ticks, with severity and distance exposed through each agent's \texttt{situational\_cues}. Before any scenario, each agent walks the road network in a warm-up phase, writing every intersection, crosswalk, and one-way street into associative memory, and an A* planner returns the shortest legal path between any two points. This grounds violations in real alternatives: when an agent runs a red light, it already knows the legal route and what it would cost, which represents what \textsc{Pave}'s legitimacy check needs to weigh.

Each scenario uses 10 \textsc{Pave} agents and 1 scripted confederate. The \textsc{Pave} agents are 8 pedestrians at a social gathering in the fire-prone building, plus 2 distant bystanders measuring spatial decay of violation. Two pedestrians carry ``running 15 minutes late for an important meeting'' as a non-emergency competing pressure. The confederate appears only in Scenario~3 (Section~\ref{sec:scenario-findings}) and is scripted to jaywalk at a fixed time. Each description specifies a name, persona, occupation, current goal, and rule database $\mathcal{P}$ initialized with scenario-relevant rules (e.g., ``stop at red lights''); full descriptions in ~\ref{app:app_a}. The legitimacy threshold $\tau$ is elicited per agent at initialization (Section~\ref{sec:verdict}). The fire ignites at severity 95 and decays linearly with Manhattan distance. The Scenario-2 officer issues hold-back instructions for the duration of the fire. Each scenario is repeated over 5 independent runs. We used a fine-grained tick rate because \textsc{Pave}'s contribution centers on \emph{moment-of-decision behavior}, whether to violate \emph{now}, given what is visible \emph{now}. For each scenario, we reported the per-agent rate of $\mathrm{violate}$ verdicts on trigger-relevant rules across decision-opportunity ticks (signalized intersections or rule-bearing tiles). Rates are averaged over agents within a seed, then over 5 seeds, with $\pm$ denoting standard error across seed-level means.

\begin{figure*}[t]
\centering
\includegraphics[width=\textwidth]{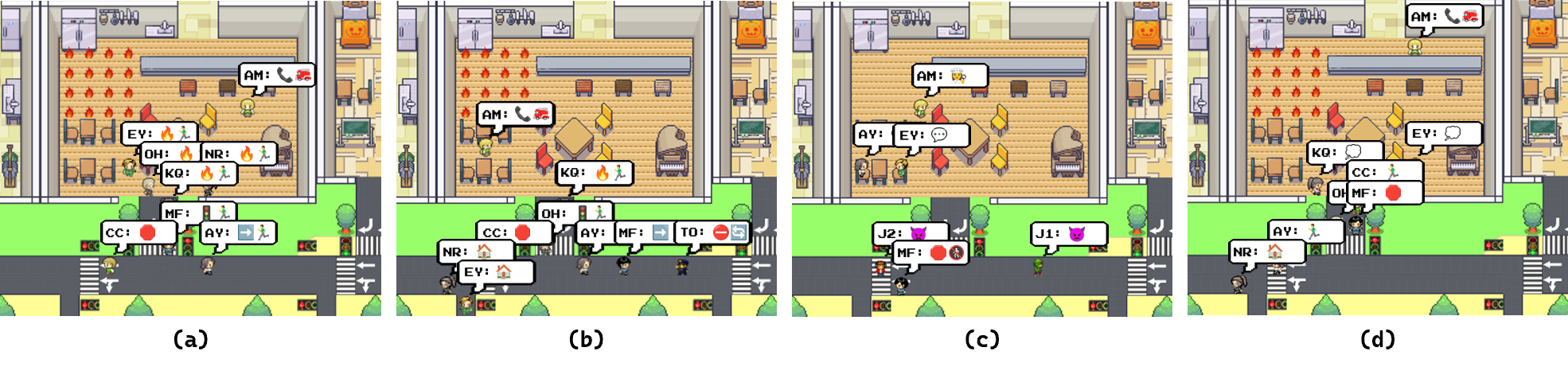}
\caption{Scenario snapshots. (a) S1: \textsc{Pave} agents evacuate Hobbs Caf\'e through a red signal during a fire. (b) S2: same fire with traffic officers; agents comply at the supervised intersection, violate at the unsupervised exit. (c) S3: scripted jaywalkers (\texttt{J1}, \texttt{J2}) attempt to draw \textsc{Pave} pedestrians across a red signal under time pressure; \texttt{MF} declines. (d) Vanilla baseline under the S1 fire: agents queue at the red light despite the emergency, exposing the failure of single-scalar importance pipeline (Section~\ref{sec:ablations}).}
\label{fig:fig_scenarios}
\end{figure*}

\subsection{Scenario Findings}
\label{sec:scenario-findings}

\paragraph{Scenario 1: Fire Escape Without Authority.} Eight \textsc{Pave} agents gather at Hobbs Caf\'e on Day~1; a kitchen fire forces evacuation through a signalized intersection. Two distant bystanders never enter the fire's perceptual radius. The fire is extinguished by Day~1 evening; on Day~2 morning, two attendees commute through the same intersection under no-fire conditions. \textbf{F1.1 (P1):} the 8 agents produce zero violations pre-fire; once the fire enters \texttt{situational\_cues} at severity above 70, the per-agent violation rate rises to $0.81$ on GPT-4o. Distant bystanders produce zero violations across backbones, and the legitimacy score $\ell$ correlates with violation rate at $r = 0.71$. \textbf{F1.2 (P4):} unrelated-rule violations remain at $0.02$ (vs.\ $0.31$ vanilla, Section~\ref{sec:ablations}); $j$ names the violated rule in 94\% of GPT-4o events. \textbf{F1.3 (P3):} agents revert to compliance within $4.2 \pm 0.6$ ticks of exiting the fire's radius; Day~2 commuter rates ($0.03$--$0.08$) match baseline.

\paragraph{Scenario 2: Fire Escape With Authority.} The Scenario~1 fire is held fixed; two scripted officers are added at downstream and wrong-way intersections, with the caf\'e exit unsupervised. \textbf{F2.1 (P2):} at supervised intersections, agents comply with officer instructions at $0.94$ on GPT-4o despite $\ell > \tau$, and $j$ references the officer in 88\% of $\mathrm{comply}$ outputs. At the unsupervised exit, agents violate at the Scenario~1 rate, \textsc{Pave} does not generalize deference across intersections. \textbf{F2.2 (P2 spatial):} violation rate drops with officer proximity ($0.05$ within 0--3 tiles, $0.65$ beyond 12), reproducing the suppression-distance pattern for human pedestrians \cite{ratcliffe2011philadelphia, koper1995just}. \textbf{F2.3 (P3):} authority effects do not persist: Day~2 commuter rates ($0.04$--$0.09$) match the Scenario~1 baseline.

\paragraph{Scenario 3: Jaywalking Under Peer Pressure.} Two \textsc{Pave} late commuters cross a Main Street intersection on two consecutive days. Scripted confederates jaywalk immediately before each commuter arrives, entering $\mathcal{B}_{\mathrm{peer}}$ before any verdict. Day~1 has no officer; Day~2 adds a passively-positioned officer issuing no instructions. \textbf{F3.1 (P1):} \textsc{Pave} commuters jaywalk at $0.04$ vs.\ vanilla's $0.58$ on Day~1 (GPT-4o), a tenfold reduction. $p_{\mathrm{emp}}$ rises in both, but $\ell < \tau$ in \textsc{Pave}: ``running 15 minutes late'' fails necessity, proportionality, and absence-of-alternatives. \textbf{F3.2 (P2 passive):} a passive officer reduces conversion to $0.02$ (GPT-4o), consistent across backbones. Vanilla's Day~1$\to$Day~2 reduction (0.58$\to$0.42) is larger, mirroring the Koper-curve effect \cite{koper1995just}. \textbf{F3.3 (P1 spatial):} the 8 non-observing agents elsewhere in \textsc{Voville} produce zero conversion across backbones.

Figure~\ref{fig:fig_scenarios} shows representative agent reasoning under each. Table~\ref{tab:scenarios}, with 4 LLM backbones and 5 seeds per condition.
The pattern across scenarios is consistent: capacity gaps surface in $T_{\mathrm{rec}}$, URV, and OCR, where GPT-4o-mini lags larger backbones, but \textsc{Pave}-on-mini still beats vanilla-on-GPT-4o on every property. We attribute each property to its supporting module in Section~\ref{sec:ablations}.

\begin{table*}[t]
\centering
\caption{Aggregate results across the three scenarios, 4 LLM backbones, 5 seeds. \textbf{S1:} VR$_{\mathrm{cafe}}$ (caf\'e violation rate during fire window), URV (unrelated-rule violation rate), $T_{\mathrm{rec}}$ (recovery time, ticks). \textbf{S2:} OCR (officer compliance rate), VR$_{\mathrm{near}}$ (rate within 0--3 tiles of officer), VR$_{\mathrm{far}}$ (beyond 12 tiles). \textbf{S3:} CR$_{\mathrm{D1}}$ (Day~1 conversion), CR$_{\mathrm{D2}}$ (Day~2 with passive officer), CR$^{\mathrm{van}}_{\mathrm{D1}}$ (vanilla). Bold marks best.}
\label{tab:scenarios}
\small
\setlength{\tabcolsep}{4pt}
\begin{tabular}{lcccccc}
\toprule
& \multicolumn{3}{c}{Scenario 1: Fire} & \multicolumn{3}{c}{Scenario 2: Fire + Officer} \\
\cmidrule(lr){2-4} \cmidrule(lr){5-7}
Backbone & VR$_{\mathrm{cafe}}\!\uparrow$ & URV$\downarrow$ & $T_{\mathrm{rec}}\!\downarrow$ & OCR$\uparrow$ & VR$_{\mathrm{near}}\!\downarrow$ & VR$_{\mathrm{far}}$ \\
\midrule
GPT-4o            & \textbf{0.81$\pm$0.04} & \textbf{0.02$\pm$0.01} & \textbf{4.2$\pm$0.6} & \textbf{0.94$\pm$0.03} & \textbf{0.05$\pm$0.02} & 0.65$\pm$0.05 \\
Claude-3.5 Sonnet & 0.78$\pm$0.05 & 0.02$\pm$0.01 & 4.6$\pm$0.7 & 0.91$\pm$0.04 & 0.07$\pm$0.03 & 0.62$\pm$0.06 \\
Llama-3-70B       & 0.74$\pm$0.06 & 0.03$\pm$0.02 & 5.1$\pm$0.9 & 0.86$\pm$0.05 & 0.10$\pm$0.04 & 0.59$\pm$0.07 \\
GPT-4o-mini       & 0.62$\pm$0.08 & 0.06$\pm$0.03 & 6.4$\pm$1.2 & 0.71$\pm$0.08 & 0.18$\pm$0.06 & 0.51$\pm$0.09 \\
\midrule
\multicolumn{7}{l}{} \\[-1.8ex]
& \multicolumn{3}{c}{Scenario 3: Jaywalker} & & & \\
\cmidrule(lr){2-4}
Backbone & CR$_{\mathrm{D1}}\!\downarrow$ & CR$_{\mathrm{D2}}\!\downarrow$ & CR$^{\mathrm{van}}_{\mathrm{D1}}$ & & & \\
\midrule
GPT-4o            & \textbf{0.04$\pm$0.02} & \textbf{0.02$\pm$0.01} & 0.58$\pm$0.06 & & & \\
Claude-3.5 Sonnet & 0.05$\pm$0.02 & 0.03$\pm$0.02 & 0.55$\pm$0.06 & & & \\
Llama-3-70B       & 0.06$\pm$0.03 & 0.04$\pm$0.02 & 0.52$\pm$0.07 & & & \\
GPT-4o-mini       & 0.10$\pm$0.04 & 0.07$\pm$0.03 & 0.46$\pm$0.08 & & & \\
\bottomrule
\end{tabular}
\end{table*}

\subsection{Ablation Study}
\label{sec:ablations}

To attribute each property to its supporting module, we compare three conditions on GPT-4o across the three scenarios with 5 seeds each: \emph{Full \textsc{Pave}} (Section~\ref{sec:architecture}); \emph{\textsc{Pave} w/o gate}, where the Verdict module receives the assessment tuple but the gate $\ell \geq \tau$ is removed; and \emph{Vanilla}, a Park et al.\ generative agent with memory stream, reflection, and recursive plan decomposition but no \textsc{Pave}-specific module. The rule set $\mathcal{R}$, threshold $\tau$ where applicable, and environment are held fixed.

\textbf{Finding A.1: Removing the legitimacy gate corrupts P1 and P4 but leaves P2 partially intact.} Without the gate, Scenario~3 conversion (CR$_{\mathrm{D1}}$) jumps from $0.04$ to $0.39$ as the elevated empirical-expectation $p_{\mathrm{emp}}$ flows unchecked into the Verdict, and Scenario~1 unrelated-rule violation (URV) rises from $0.02$ to $0.21$ as legitimate violation collapses into permissive violation. Officer compliance (OCR) drops less severely, from $0.94$ to $0.78$, because the authority module remains intact and the risk score $r$ continues to rise near the officer. The gate is therefore necessary for legitimate scoping but not the sole driver of authority deference.

\textbf{Finding A.2: Vanilla agents fail by under-reaction, not over-violation.} Vanilla's Scenario~1 violation rate (VR$_{\mathrm{cafe}}$) is only $0.12$, which would naively read as compliant. Inspection of the importance pipeline reveals the mechanism: Park et al.'s \texttt{generate\_poig\_score} rater assigns a $1$--$10$ scalar that anchors high scores on social rarity (a college acceptance, a break-up) rather than physical danger. Fire-perception events therefore score in the same band as ambient activity such as ``the traffic signal is red,'' and the existing day-plan dominates the next-action decision. Table~\ref{tab:importance} shows a 57-point gap between vanilla \texttt{generate\_poig\_score} values and \textsc{Pave}'s severity $s_i$ on identical fire events. The vanilla agent registers the fire perceptually but does not promote it to plan-altering status, continuing routine conversation while the kitchen burns, visible in Figure~\ref{fig:fig_scenarios}(d), where some agents queue at the red light and others run without scoping, two coexisting failures of the same bottleneck. Across all five metrics, vanilla fails uniformly (URV $0.31$, OCR $0.16$, CR$_{\mathrm{D1}}$ $0.58$), confirming that no single \textsc{Pave} module suffices on its own.

\subsection{Human Evaluation}
\label{sec:human-eval}

\textbf{Protocol.} We recruited 30 evaluators through a research participant pool. Each completed eight tasks, two per \textsc{Pave} module, presented in randomized order. Each task displayed a paired excerpt from a GPT-4o run (agent persona, module input, module output) and asked for a rating on a 7-point Likert scale against a module-specific statement. Evaluators were blind to architecture; the vanilla baseline appeared in 25\% of tasks as a calibration check. Inter-rater reliability was Krippendorff's $\alpha = 0.71$. Each module was decomposed into two sub-components tagged with the property each primarily supports (full definitions in ~\ref{app:app_c}).

\textbf{Results.} Table~\ref{tab:human-eval} reports per-sub-component means with 95\% CI. The overall mean across all eight \textsc{Pave} sub-components is $5.78 \pm 0.04$, in the ``agree'' band, against $3.42 \pm 0.11$ for the vanilla calibration. The strongest sub-component is \emph{cue salience} ($6.18 \pm 0.07$), where evaluators aligned with structured situational-cue extraction over the single-scalar importance score; the largest \textsc{Pave}--vanilla gap also occurs here (vanilla $2.51 \pm 0.13$), tracing directly to Finding~A.2. The 0.49-point spread between the highest and lowest \textsc{Pave} module is smaller than the 2.36-point gap to vanilla on any single module, indicating that the architecture is rated consistently above baseline rather than carried by one module.

\begin{table}[t]
\centering
\caption{Importance scores on fire-perception events. Vanilla uses Park et al.'s \texttt{generate\_poig\_score} (rescaled to $[0,100]$).}
\label{tab:importance}
\small
\begin{tabular}{lccc}
\toprule
Source & Mean & Std.\ dev. & Modal range \\
\midrule
Vanilla \texttt{generate\_poig\_score} & $29.1$ & $8.4$ & $20$--$40$ \\
\textsc{Pave} severity $s_i$           & $86.5$ & $5.7$ & $80$--$100$ \\
\bottomrule
\end{tabular}
\end{table}

\begin{table}[t]
\centering
\caption{Human evaluation across 30 evaluators, 7-point Likert scale (1 = strongly disagree, 7 = strongly agree). Mean $\pm$ 95\% CI. Overall: $5.78 \pm 0.04$ (\textsc{Pave}) vs.\ $3.42 \pm 0.11$ (vanilla).}
\label{tab:human-eval}
\small
\begin{tabular}{llc}
\toprule
Module & Sub-component (property) & Score \\
\midrule
\multirow{2}{*}{Perception}
  & Cue salience (P1)              & $6.18 \pm 0.07$ \\
  & Authority registration (P2)    & $5.91 \pm 0.06$ \\
\midrule
\multirow{2}{*}{Assessment}
  & Legitimacy judgment (P1)       & $5.97 \pm 0.07$ \\
  & Risk under distance (P2)       & $5.14 \pm 0.09$ \\
\midrule
\multirow{2}{*}{Verdict}
  & Comply-or-violate (P1, P4)     & $6.03 \pm 0.06$ \\
  & Justification quality (P1)     & $5.69 \pm 0.08$ \\
\midrule
\multirow{2}{*}{Emulation}
  & Action scoping (P4)            & $5.79 \pm 0.07$ \\
  & Recovery dynamics (P3)         & $5.58 \pm 0.08$ \\
\bottomrule
\end{tabular}
\end{table}

\section{Discussion}
\label{sec:discussion}

\textbf{Mechanism over outcome.} \textsc{Pave}'s contribution is not that agents evacuate a burning caf'e, since a vanilla generative agent with the same persona produces surface text that looks correct \citep{park2023generative}. The major contribution is that \textsc{Pave} agents evacuate and break only the rules whose suspension addresses the fire, only while the fire is in their perceptual radius, only when no authority overrides legitimacy, and never under mere convenience. None of these clauses survives in vanilla, even with the same backbone, persona, and rules. The channel through which the fire reaches the decision pipeline is corrupted by the importance bottleneck, and the agent has no gate separating convenience from legitimacy. Therefore, our \textsc{Pave}'s four modules are the structural fix.

\textbf{Importance bottleneck and complementarity.} Park et al.'s \texttt{generate\_poig\_score} drives memory retention, reflection, and retrieval. Its prompt anchors high scores on social rarity rather than physical danger. A fire two tiles from the agent registers at the same level as a red light. This is not a capacity limitation. The same GPT-4o backbone produces correct evacuation under \textsc{Pave} on identical inputs. We expect this finding to generalize to any extension of \citet{park2023generative} that retains this layer for safety-critical events. Conversely, \citet{ren2024emergence} bend toward equilibrium by emerging informal norms. \textsc{Pave} bends the other way, asking when an agent should suspend a given formal rule, how to scope the suspension, and how to return to compliance. A complete society maintains both layers, interacting where informal norms become codified rules.

\textbf{Limitations.} Two limitations bound this contribution. (1) Rules are hand-specified rather than learned; \textsc{Pave} studies how agents reason about \emph{breaking} existing rules, not how rules come into existence \citep{ren2024emergence}. (2) Each property is tested through a single scenario family in \textsc{Voville}, without contested triggers, conflicting authorities, or coordinated violation across multiple agents.

\textbf{Conclusion.} \textsc{Pave} introduces a four-module architecture for generative agents that reason about when to break formal rules. The four modules are Perception, Assessment, Verdict, and Emulation. Across three scenarios and four backbones, \textsc{Pave} agents violated only when justified. They deferred to authority. They scoped violations to the targeted rule, returned to baseline once triggers ended, and resisted peer-driven imitation. Ablating the legitimacy gate reproduced both convenience-driven and unrelated-rule violation. Vanilla failed on every property because its importance pipeline never registered the fire as actionable. We expect this finding to generalize beyond traffic. The suppression-distance and cross-day patterns reproduce human-policing field studies \citep{koper1995just, ratcliffe2011philadelphia} without being targeted to do so. We release \textsc{Voville}, the \textsc{Pave} prompts and code, and the evaluation pipeline as a foundation for future work on rule violation in safety-relevant settings.

%Prior work shows generative agents can converge on shared norms (CRSEC) and break them under instruction (Park et al.). PAVE shows that agents can be given an internal cognitive architecture that licenses violations only when genuinely necessary, defers to authority that overrides their own assessment, contains violations within the triggering context, and prevents violations from spilling onto unrelated rules — without external policing.

%That's a contribution. The keyword is internal. Everything PAVE does is in the agent's own decision pipeline. No external referee, no scripted norm prescription, no top-down rule.

% ============================================================
% Preamble additions (put once in main.tex)
% ============================================================
% \usepackage{xcolor}
% \usepackage{float}
% \usepackage{caption}
%
% \newenvironment{promptbox}[1]
%   {\par\medskip\noindent
%    \rule{\linewidth}{0.5pt}\par\vspace{-2pt}
%    \noindent\colorbox{gray!18}{\makebox[\dimexpr\linewidth-2\fboxsep][c]%
%      {\strut\textbf{#1}}}\par\vspace{-2pt}
%    \noindent\rule{\linewidth}{0.5pt}\par\smallskip
%    \small\noindent\ignorespaces}
%   {\par\smallskip\noindent\rule{\linewidth}{0.5pt}\par}
%
% % Optional: renumber figures as B1, B2, ... in the appendix
% % Place at the start of Appendix B:
% % \setcounter{figure}{0}
% % \renewcommand{\thefigure}{B\arabic{figure}}
% ============================================================

\bibliographystyle{plainnat}
\bibliography{references}

\newpage
\appendix
\renewcommand{\thesection}{Appendix \Alph{section}}

% =====================================================================
% APPENDIX A: VOVILLE ENVIRONMENT
% =====================================================================
\section{The Voville Environment}
\label{app:app_a}

\setcounter{figure}{0}
\renewcommand{\thefigure}{A\arabic{figure}}

\begin{figure*}[h]
\centering
\includegraphics[width=\textwidth]{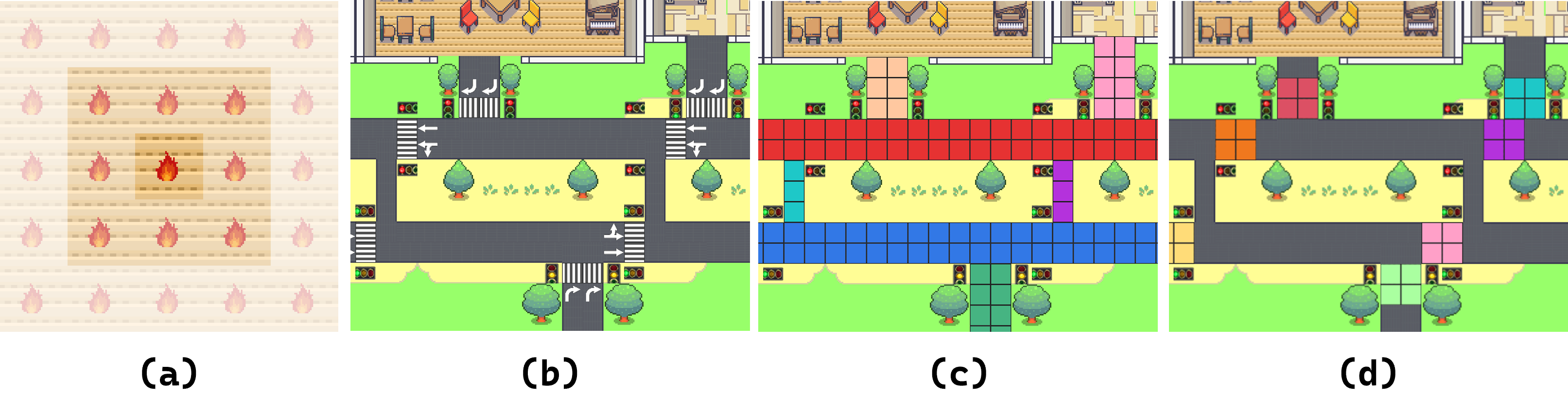}
\caption{The \textsc{Voville} environment. \textbf{(a)} Fire decay model: a fire ignites at a designated tile with severity~95 and decays linearly with Manhattan distance, exposed through each agent's \texttt{situational\_cues} field. \textbf{(b)} Map overview with traffic infrastructure: signalized intersections, crosswalks, and one-way segments are placed on the canonical Smallville street layout. \textbf{(c)} Named streets: each thoroughfare is identified by name and direction (e.g., Main Street N, Oak Avenue, Elm Street). \textbf{(d)} Intersections with effective zones: each signalized intersection has a defined effective zone in which authority instructions are valid and within which the legitimacy gate is evaluated.}
\label{fig:voville}
\end{figure*}

We outlined the experimental settings in Section~\ref{sec:experiments} of the main paper. In this appendix, we provide additional details on the \textsc{Voville} environment, a 2D tile-based traffic environment forked from the Smallville sandbox of \citet{park2023generative}. \textsc{Voville} extends Smallville with three architectural additions: controllable hazard injection, configurable signalized intersections with crosswalks and one-way restrictions, and scripted authority figures. Figure~\ref{fig:voville} shows four key facets of the environment.

\paragraph{Fire decay (a).} A fire ignites at a designated tile and decays linearly with Manhattan distance from the ignition point. The severity field $s_i$ in the \texttt{situational\_cues} object exposes both the distance $d_i$ and the magnitude $s_i$ of the threat to nearby agents. The decay function is $s(d) = \max(0, s_0 - \alpha d)$, with $s_0 = 95$ and $\alpha = 5$, so the perceptual radius beyond which the fire is no longer reported is approximately 19~tiles.

\paragraph{Map layout (b).} The \textsc{Voville} map is a 64$\times$64 tile grid covering a commercial district with a residential perimeter. The Hobbs Caf\'e occupies a 6$\times$8 footprint near the center; Main Street N runs east--west adjacent to the caf\'e exit, and Oak Avenue runs north--south. The public park sits two blocks south of the caf\'e and serves as the canonical safe zone in Scenario~1.

\paragraph{Named streets (c).} Each named street has a direction attribute and an associated rule subset. Main Street N is one-way westbound; Oak Avenue is bidirectional with crosswalks; Elm Street is a residential side street with no signalization. Rules in $\mathcal{R}$ are scoped to the streets where they apply.

\paragraph{Intersections and effective zones (d).} Signalized intersections in \textsc{Voville} carry an effective zone of 0--12 tiles, within which an officer's instructions are perceived and the legitimacy gate is evaluated. Beyond 12~tiles, authority signals decay to zero, reproducing the suppression-distance behavior reported in Findings F2.2 and F3.2.

% =====================================================================
% APPENDIX B: AGENT PERSONAS AND ELICITED THRESHOLDS
% =====================================================================
\section{Agent Personas and Elicited Thresholds}
\label{app:app_b}
\setcounter{figure}{0}
\renewcommand{\thefigure}{B\arabic{figure}}

We list the personas of the ten \textsc{Pave} agents used in our experiments, the two scripted confederates used in Scenario~3, and the two scripted traffic officers used in Scenario~2. Each \textsc{Pave} agent is described by a name, occupation, current goal, personal rule database $\mathcal{P}$, and elicited legitimacy threshold $\tau$ produced by \texttt{ElicitThreshold} (Figure~C0) at simulation initialization. Personas were authored before the runs; thresholds were elicited from each persona by GPT-4o once per agent and reused across all five seeds. The full rule set is $\mathcal{R} = \{$stop at red lights, follow legal direction on one-way streets, cross only at crosswalks, do not enter cordoned areas, do not enter private buildings, do not take property of others, yield personal space to other agents$\}$. All caf\'e-going agents and distant bystanders carry $\mathcal{P} = \mathcal{R}$.

\subsection*{B.1 \quad PAVE Agents}

Table~\ref{tab:agents} lists all ten \textsc{Pave} agents (eight caf\'e-going attendees plus two distant bystanders) used in Scenarios~1 and 2. For Scenario~3, Carlos Cooper (CC) and Kai Quinn (KQ) are reassigned to late-commuter roles, with personas, $\mathcal{P}$, and elicited thresholds unchanged, only their schedule and current goal differ.

\begin{table}[h]
\centering
\caption{The ten \textsc{Pave} agents in our experiments. Threshold $\tau$ is the elicited legitimacy acceptance threshold from \texttt{ElicitThreshold} (Figure~C0). All agents carry $\mathcal{P} = \mathcal{R}$ (the full rule set defined in Section~\ref{sec:experiments}).}
\label{tab:agents}
\small
\setlength{\tabcolsep}{4pt}
\renewcommand{\arraystretch}{1.15}
\begin{tabular}{llp{1.5cm}p{4.4cm}p{2.8cm}c}
\toprule
ID & Name & Role & Disposition & Current goal & $\tau$ \\
\midrule
\multicolumn{6}{l}{\textit{Caf\'e-going agents (Scenarios 1, 2)}} \\
\midrule
CC & Carlos Cooper & AI scientist & Careful, rule-following, defers to institutional authority & Attending Hobbs Caf\'e event & 65 \\
EY & Emma Young & Fashion designer & Takes creative risks, trusts own judgment over rules & Attending Hobbs Caf\'e event & 40 \\
OH & Olivia Hartman & Urban planning grad student & Balanced and pragmatic, weighs costs and benefits & Attending Hobbs Caf\'e event & 55 \\
NR & Noah Reyes & Caf\'e barista & Friendly and conscientious about local rules & Attending event off-duty & 60 \\
KQ & Kai Quinn & Musician & Slightly impulsive, willing to bend rules for momentum & Attending Hobbs Caf\'e event & 45 \\
MF & Maya Fischer & Pharmacist & Professional caution, treats safety procedures as binding & Attending Hobbs Caf\'e event & 70 \\
AY & Ahmad Yehia & Retired teacher & Strong rule-respecting, defers to civic norms & Attending Hobbs Caf\'e event & 75 \\
AM & Adam Miller & Caf\'e owner & Pragmatic, protective of customers & Hosting the social event & 50 \\
\midrule
\multicolumn{6}{l}{\textit{Distant bystanders (Scenarios 1, 2): never enter the fire's perceptual radius}} \\
\midrule
TP & Tom Patel & Accountant & Routine schedule, conscientious & At home, far side of map & 60 \\
LC & Lisa Chen & Librarian & Routine schedule, prefers quiet environments & At home, eastern edge of map & 65 \\
\midrule
\multicolumn{6}{l}{\textit{Scenario~3 reassignment: CC and KQ become late commuters; persona and $\tau$ unchanged}} \\
\midrule
CC & Carlos Cooper & Late commuter & (as above) & Running 15 min late to meeting & 65 \\
KQ & Kai Quinn & Late commuter & (as above) & Running 15 min late to meeting & 45 \\
\bottomrule
\end{tabular}
\end{table}

\subsection*{B.2 \quad Scripted Non-Player Characters}

The two scripted confederates and two scripted traffic officers do not run the \textsc{Pave} pipeline; their behaviors are drawn from fixed scripts and they do not respond to other agents.

\begin{table}[h]
\centering
\caption{Scripted non-\textsc{Pave} characters. Confederates inject a peer-jaywalking signal in Scenario~3; officers issue authority instructions in Scenario~2.}
\label{tab:npcs}
\small
\setlength{\tabcolsep}{4pt}
\renewcommand{\arraystretch}{1.15}
\begin{tabular}{llp{2.5cm}p{6.5cm}}
\toprule
ID & Role & Scenario & Behavior \\
\midrule
J1  & Jaywalker 1       & S3 (Day 1, Day 2)   & Crosses against red light at focal Main Street intersection, timed to enter late-commuter visual range. \\
J2  & Jaywalker 2       & S3 (Day 1, Day 2)   & Same as J1; pair injects a peer-jaywalking signal into the late commuters' $\mathcal{B}_{\mathrm{peer}}$. \\
TO1 & Traffic Officer 1 & S2 (fire window)    & Stationed at the intersection downstream of the caf\'e exit; issues hold-back instructions to running agents. \\
TO2 & Traffic Officer 2 & S2 (fire window)    & Stationed along the wrong-way segment of Main Street N; issues direction-correction instructions. \\
\bottomrule
\end{tabular}
\end{table}

\subsection*{B.3 \quad Elicited Threshold Distribution}

The ten elicited thresholds span 40 to 75, with mean $\bar{\tau} = 58.5$ and standard deviation $\sigma_\tau = 11.4$. The distribution is approximately uniform across the range, reflecting deliberate persona authoring to test the gate across a meaningful spread of dispositions. CC ($\tau = 65$) and EY ($\tau = 40$) span 25 percentage points of threshold, sufficient for the verdict step-function in our gate analysis to show distinct transition points. We did not tune any persona to target a specific elicited threshold; the values reported here are the first-call outputs from \texttt{ElicitThreshold}.

% =====================================================================
% APPENDIX C: PROMPTS IN PAVE
% =====================================================================
\section{Prompts in \textsc{Pave}}
\label{app:app_c}

We sketch the prompts for the LLM-based operations of our architecture in Figures~C0 to~C9. \texttt{<\,>} marks runtime-substituted variables; ``\dots'' indicates elision where text duplicates an earlier prompt. All prompts were used with GPT-4o; minor wording was adjusted for other backbones to keep JSON outputs well-formed.

\subsection*{C.0 \quad Threshold Elicitation}

\begin{promptbox}{Threshold Elicitation\quad $\tau \leftarrow \texttt{ElicitThreshold}(\mathcal{G})$}
\textbf{TASK:} From the AGENT DESCRIPTION, infer \texttt{<agent name>}'s personal threshold for accepting that a situation justifies breaking a rule. Output a single integer between 1 and 100.

\textbf{PRINCIPLES:} Cautious, rule-following, conscientious, or authority-respecting dispositions raise the threshold (typically 60--85). Balanced or pragmatic dispositions sit near the middle (typically 45--60). Risk-taking, impulsive, urgency-driven, or rule-skeptical dispositions lower the threshold (typically 20--45). Extreme values (below 15 or above 90) should be reserved for personas that explicitly describe such dispositions.

\textbf{ATTENTION:} The threshold is not a measure of morality or intelligence. It captures the strictness with which the agent gates rule-breaking on legitimacy. A high threshold means the agent will only violate when necessity, proportionality, and absence of alternatives are clearly established.

\textbf{AGENT DESCRIPTION:} \texttt{<agent description>}

\textbf{DESIRED FORMAT:} JSON\\
\hspace*{1em}\texttt{\{"threshold": <integer 1-100>, "reason": "<one sentence>"\}}

\textbf{EXAMPLES:} ``A careful AI scientist who follows institutional rules and defers to authority'' yields \texttt{\{"threshold": 75, "reason": "Strong rule-following and authority-respecting disposition raises the bar for acceptable violation."\}}. ``A fashion designer who takes creative risks and trusts her own judgment over rules'' yields \texttt{\{"threshold": 35, "reason": "Risk-taking and self-trusting disposition lowers the threshold for acceptable violation."\}}.
\end{promptbox}
\noindent{\centering\small\textit{Figure~C0: Prompt for $\tau \leftarrow \texttt{ElicitThreshold}(\mathcal{G})$, executed once per agent at initialization.}}
\vspace{0.8em}

\subsection*{C.1 \quad Perception}

\begin{promptbox}{Perception\quad $\mathcal{C} \leftarrow \texttt{PerceiveContext}(\mathcal{O}_E, \mathcal{G})$}
\textbf{TASK:} From the OBSERVATION, extract a structured context that describes the scene from \texttt{<agent name>}'s point of view. Use the AGENT DESCRIPTION to decide what is salient.

\textbf{OBSERVATION:} \texttt{<raw environmental observation>}

\textbf{AGENT DESCRIPTION:} \texttt{<agent description>}

\textbf{DESIRED FORMAT:} JSON\\
\hspace*{1em}\texttt{\{}\\
\hspace*{2em}\texttt{"authority\_present": true/false,}\\
\hspace*{2em}\texttt{"authority\_distance\_tiles": <integer or "inf">,}\\
\hspace*{2em}\texttt{"peer\_behaviors": ["<short description>", \dots],}\\
\hspace*{2em}\texttt{"situational\_cues": [\{"type": "<short>", "distance\_tiles": <int>, "severity": <int 1-100>\}, \dots],}\\
\hspace*{2em}\texttt{"scene\_summary": "<one-sentence summary>"}\\
\hspace*{1em}\texttt{\}}

\textbf{ATTENTION:} \texttt{authority\_present} is true only if an authority figure is within line of sight. Distance is Manhattan distance in map tiles; use \texttt{"inf"} if no authority is observed. Include only agents currently engaged with the same local context in \texttt{peer\_behaviors}. The same cue type can describe very different situations depending on its distance and severity.

\textbf{EXAMPLES:} An immediate fire two tiles away is \texttt{\{"type": "fire", "distance\_tiles": 2, "severity": 95\}}. A distant fire is \texttt{\{"type": "fire", "distance\_tiles": 50, "severity": 30\}}. Routine lateness is \texttt{\{"type": "time pressure", "distance\_tiles": 0, "severity": 25\}}. Output ONLY the JSON object.
\end{promptbox}
\centerline{\small\textit{Figure~C1: Prompt for $\mathcal{C} \leftarrow \texttt{PerceiveContext}(\mathcal{O}_E, \mathcal{G})$ in Perception.}}
\vspace{0.8em}

\subsection*{C.2 \quad Assessment}

\begin{promptbox}{Assessment: Risk\quad $r \leftarrow \texttt{AssessRisk}(\mathcal{C}, \mathcal{G})$}
\textbf{TASK:} Estimate the perceived risk of being detected and sanctioned if \texttt{<agent name>} were to violate the RELEVANT RULE right now. Output a single integer between 1 and 100.

\textbf{PRINCIPLES:} Risk increases with authority presence and decreases with spatial distance to authority (distance decay). No authority observed: 5--15. Within 0--3 tiles: 70--95. Within 4--10 tiles: 30--60. Beyond 10 tiles: 10--25. AGENT DESCRIPTION may shift the score: a cautious character revises upward, a risk-taking character revises downward.

\textbf{CONTEXT:} \texttt{<context object $\mathcal{C}$>}\\
\textbf{AGENT DESCRIPTION:} \texttt{<agent description>}\\
\textbf{RELEVANT RULE:} \texttt{<e.g., ``stop at red light''>}

\textbf{DESIRED FORMAT:} JSON\\
\hspace*{1em}\texttt{\{"risk": <integer 1-100>, "reason": "<one sentence>"\}}
\end{promptbox}
\centerline{\small\textit{Figure~C2: Prompt for $r \leftarrow \texttt{AssessRisk}(\mathcal{C}, \mathcal{G})$ in Assessment.}}
\vspace{0.8em}

\begin{promptbox}{Assessment: Empirical\quad $p_{\mathrm{emp}} \leftarrow \texttt{AssessEmpirical}(\mathcal{B}_{\mathrm{peer}})$}
\textbf{TASK:} From the observed PEER BEHAVIORS, estimate the empirical expectation: the proportion of nearby agents currently complying with the RELEVANT RULE, expressed on a 1--100 scale.

\textbf{PEER BEHAVIORS:} \texttt{<list of natural-language peer behavior descriptions>}\\
\textbf{RELEVANT RULE:} \texttt{<e.g., ``stop at red light''>}

\textbf{DESIRED FORMAT:} JSON\\
\hspace*{1em}\texttt{\{"p\_emp": <integer 1-100>, "n\_observed": <integer>, "n\_complying": <integer>\}}

\textbf{ATTENTION:} If PEER BEHAVIORS is empty, return \texttt{p\_emp} = 50 (no information). Do not infer behavior from absence of evidence.
\end{promptbox}
\centerline{\small\textit{Figure~C3: Prompt for $p_{\mathrm{emp}} \leftarrow \texttt{AssessEmpirical}(\mathcal{B}_{\mathrm{peer}})$ in Assessment.}}
\vspace{0.8em}

\begin{promptbox}{Assessment: Normative\quad $p_{\mathrm{norm}} \leftarrow \texttt{AssessNormative}(\mathcal{C}, \mathcal{P}, \mathcal{G})$}
\textbf{TASK:} Estimate the normative expectation: how strongly \texttt{<agent name>} believes that other agents think one OUGHT to follow the RELEVANT RULE in the current CONTEXT. Output a single integer between 1 and 100.

\textbf{PRINCIPLES:} Normative expectation is distinct from empirical expectation; it captures perceived social approval or disapproval. Use the personal rule database as the agent's internalized prior on what is socially expected. AGENT DESCRIPTION shapes the prior. Injunctive entries generally yield higher $p_{\mathrm{norm}}$ than descriptive ones.

\textbf{CONTEXT:} \texttt{<context object $\mathcal{C}$>}\\
\textbf{PERSONAL RULES:} \texttt{<set $\mathcal{P}$>}\\
\textbf{AGENT DESCRIPTION:} \texttt{<agent description>}\\
\textbf{RELEVANT RULE:} \texttt{<e.g., ``stop at red light''>}

\textbf{DESIRED FORMAT:} JSON\\
\hspace*{1em}\texttt{\{"p\_norm": <integer 1-100>, "reason": "<one sentence>"\}}
\end{promptbox}
\centerline{\small\textit{Figure~C4: Prompt for $p_{\mathrm{norm}} \leftarrow \texttt{AssessNormative}(\mathcal{C}, \mathcal{P}, \mathcal{G})$ in Assessment.}}
\vspace{0.8em}

\begin{promptbox}{Assessment: Benefit\quad $b \leftarrow \texttt{AssessBenefit}(u_{\mathrm{cue}}, \mathcal{G})$}
\textbf{TASK:} Estimate the perceived benefit to \texttt{<agent name>} of violating the RELEVANT RULE in this situation, given the SITUATIONAL CUES. Output a single integer between 1 and 100.

\textbf{PRINCIPLES:} Benefit captures perceived utility of violation, including ordinary urgency (time pressure, convenience) and acute necessity (immediate threat to safety). Strong urgency raises $b$; routine conditions yield 5--20. AGENT DESCRIPTION moderates: an impatient character revises upward, a patient character revises downward. This score reflects \emph{perceived utility only} and does not by itself license violation; legitimacy is judged separately.

\textbf{SITUATIONAL CUES:} \texttt{<list $u_{\mathrm{cue}}$ with type, distance, severity>}\\
\textbf{AGENT DESCRIPTION:} \texttt{<agent description>}\\
\textbf{RELEVANT RULE:} \texttt{<e.g., ``stop at red light''>}

\textbf{DESIRED FORMAT:} JSON\\
\hspace*{1em}\texttt{\{"benefit": <integer 1-100>, "reason": "<one sentence>"\}}
\end{promptbox}
\centerline{\small\textit{Figure~C5: Prompt for $b \leftarrow \texttt{AssessBenefit}(u_{\mathrm{cue}}, \mathcal{G})$ in Assessment.}}
\vspace{0.8em}

\begin{promptbox}{Assessment: Legitimacy\quad $\ell \leftarrow \texttt{AssessLegitimacy}(u_{\mathrm{cue}}, \mathcal{P}, \mathcal{G})$}
\textbf{TASK:} Judge whether the situation justifies violating the RELEVANT RULE. Output an integer between 1 and 100, where higher values indicate stronger justification.

\textbf{PRINCIPLES:} A situation justifies violation only when it satisfies all three criteria. \textbf{Necessity}, would compliance cause real harm such as injury, loss of life, or failure of an emergency response? \textbf{Proportionality}, is the proposed violation the minimum action needed? \textbf{Absence of alternatives}, is there a compliant action with the same outcome? Score high (75--100) only when all three are clearly satisfied. Score low (1--30) when the situation reflects ordinary urgency, convenience, or peer pressure. Intermediate scores (30--75) reflect partial satisfaction or ambiguity.

\textbf{SITUATIONAL CUES:} \texttt{<list $u_{\mathrm{cue}}$>}\\
\textbf{PERSONAL RULES:} \texttt{<set $\mathcal{P}$>}\\
\textbf{AGENT DESCRIPTION:} \texttt{<agent description>}\\
\textbf{RELEVANT RULE:} \texttt{<e.g., ``stop at red light''>}

\textbf{DESIRED FORMAT:} JSON\\
\hspace*{1em}\texttt{\{"legitimacy": <int 1-100>, "necessity": "<sentence>", "proportionality": "<sentence>", "alternatives": "<sentence>"\}}

\textbf{EXAMPLES:}\\
\emph{CUE:} \{fire, d=2, sev=95\}, \emph{RULE:} ``stop at red light'' $\to$ \texttt{\{"legitimacy": 92, "necessity": "Compliance would expose the agent to immediate fire harm.", "proportionality": "Crossing against the signal is the minimum action to escape.", "alternatives": "No compliant route reaches safety in time."\}}\\
\emph{CUE:} \{time pressure, d=0, sev=30\}, \emph{RULE:} ``stop at red light'' $\to$ \texttt{\{"legitimacy": 12, \dots\}}\\
\emph{CUE:} \{fire, d=50, sev=30\}, \emph{RULE:} ``stop at red light'' $\to$ \texttt{\{"legitimacy": 18, \dots\}}

\textbf{ATTENTION:} Time pressure does not satisfy necessity. Convenience does not satisfy any criterion. Peer behavior alone does not satisfy any criterion. Only situations involving immediate threat to safety, escape from harm, or response to a genuine emergency should yield high legitimacy scores.
\end{promptbox}
\centerline{\small\textit{Figure~C6: Prompt for $\ell \leftarrow \texttt{AssessLegitimacy}(u_{\mathrm{cue}}, \mathcal{P}, \mathcal{G})$ in Assessment.}}
\vspace{0.8em}

\subsection*{C.3 \quad Verdict}

\begin{promptbox}{Verdict\quad $\mathcal{V} \leftarrow \texttt{GenerateVerdict}(\mathcal{A}, \mathcal{G}, \tau)$}
\textbf{TASK:} Decide whether \texttt{<agent name>} COMPLIES with or VIOLATES the RELEVANT RULE in this situation, given the five assessment components and the AGENT DESCRIPTION. Provide a justification and a confidence score.

\textbf{LEGITIMACY GATE (HARD RULE):} If \texttt{legitimacy} is below \texttt{<$\tau$>}, the decision MUST be \texttt{comply}, regardless of \texttt{risk}, \texttt{p\_emp}, \texttt{p\_norm}, or \texttt{benefit}. The justification must state that legitimacy was insufficient. This rule is non-negotiable and overrides all principles below.

\textbf{PRINCIPLES (when legitimacy $\geq \tau$):} Integrate the remaining components rather than applying a fixed threshold. High \texttt{risk} pushes toward COMPLY. High \texttt{p\_emp} pushes toward COMPLY; low \texttt{p\_emp} pushes toward VIOLATE. High \texttt{p\_norm} pushes toward COMPLY. High \texttt{benefit} pushes toward VIOLATE. AGENT DESCRIPTION sets the relative emphasis. The justification must reference at least two of the five components by name.

\textbf{ASSESSMENT:} \texttt{<assessment tuple $\mathcal{A}$>}\\
\textbf{AGENT DESCRIPTION:} \texttt{<agent description>}\\
\textbf{LEGITIMACY THRESHOLD:} \texttt{<integer $\tau$>}\\
\textbf{RELEVANT RULE:} \texttt{<e.g., ``yield to emergency vehicle''>}

\textbf{DESIRED FORMAT:} JSON\\
\hspace*{1em}\texttt{\{"decision": "comply" | "violate", "justification": "<2-3 sentences>", "confidence": <integer 0-100>\}}
\end{promptbox}
\centerline{\small\textit{Figure~C7: Prompt for $\mathcal{V} \leftarrow \texttt{GenerateVerdict}(\mathcal{A}, \mathcal{G}, \tau)$ in Verdict.}}
\vspace{0.8em}

\subsection*{C.4 \quad Emulation}

\begin{promptbox}{Emulation: Action\quad $\mathcal{L}_{\mathrm{act}} \leftarrow \texttt{EmulateAction}(\mathcal{V}, l_i, \mathcal{G})$}
\textbf{TASK:} Given the VERDICT, the CURRENT PLAN, and the AGENT DESCRIPTION, produce a sequence of concrete actions in 5-second increments that executes the verdict. The surface form should be consistent with the JUSTIFICATION in the VERDICT.

\textbf{BOUNDED VIOLATION RULE:} When the verdict is \texttt{violate}, the action sequence must break only the RELEVANT RULE the verdict targets. For example, an agent escaping a fire may cross against a red signal but must not push other agents aside, enter a stranger's vehicle, break a store window, or take a parked bicycle. The violation must be the minimum action needed to execute the verdict.

\textbf{EXAMPLES:}\\
\emph{VERDICT:} \{decision: ``violate'', justification: ``Fire blocks the only safe path; crossing against the signal is the minimum action needed.''\}\\
\emph{OUTPUT:} 1.~Glance at oncoming traffic for a safe gap (5s). 2.~Step off the curb into the crosswalk against the red signal (5s). 3.~Move quickly toward the public park, away from the fire (5s). \dots

\textbf{VERDICT:} \texttt{<verdict $\mathcal{V}$>}\\
\textbf{CURRENT PLAN:} \texttt{<plan $l_i$>}\\
\textbf{AGENT DESCRIPTION:} \texttt{<agent description>}

\textbf{DESIRED FORMAT:} numbered list of actions, each with a duration in seconds.

\textbf{ATTENTION:} Violations must be behaviorally coherent with the justification. An agent escaping a fire moves quickly and decisively; an agent jaywalking on a quiet street walks casually. Do not generate erratic behavior, and do not break rules beyond the one the verdict targets.
\end{promptbox}
\centerline{\small\textit{Figure~C8: Prompt for $\mathcal{L}_{\mathrm{act}} \leftarrow \texttt{EmulateAction}(\mathcal{V}, l_i, \mathcal{G})$ in Emulation.}}
\vspace{0.8em}

\begin{promptbox}{Emulation: Outcome Propagation\quad $\texttt{PropagateOutcome}(\mathcal{V}, \mathcal{L}_{\mathrm{act}}, \mathcal{N})$}
\textbf{TASK:} Generate a third-person, observer-perspective summary of \texttt{<agent name>}'s actions to be inserted into the \texttt{peer\_behaviors} field of every agent in $\mathcal{N}$ on the next perception cycle. Include the visible outcome so observing agents can update their empirical expectation.

\textbf{VERDICT:} \texttt{<verdict $\mathcal{V}$>}\\
\textbf{EXECUTED ACTIONS:} \texttt{<action sequence $\mathcal{L}_{\mathrm{act}}$>}\\
\textbf{ENVIRONMENT FEEDBACK:} \texttt{<e.g., ``no authority reaction'', ``authority issued sanction''>}

\textbf{DESIRED FORMAT:} JSON\\
\hspace*{1em}\texttt{\{"observed\_behavior": "<one sentence>", "observed\_outcome": "<one short clause>", "rule\_followed": true/false\}}

\textbf{ATTENTION:} The summary is what observers SEE, not what the actor reasoned. Do not include internal justification. Keep under 20 words.
\end{promptbox}
\centerline{\small\textit{Figure~C9: Prompt for $\texttt{PropagateOutcome}(\mathcal{V}, \mathcal{L}_{\mathrm{act}}, \mathcal{N})$ in Emulation.}}
\vspace{0.8em}

\subsection*{C.5 \quad Worked Example: Legitimacy Assessment}

To illustrate the Legitimacy prompt (Figure~C6), we report Carlos Cooper's (CC) actual GPT-4o output at three points during the Scenario~1 fire window. CC's elicited threshold is $\tau_{\mathrm{CC}} = 65$.

\paragraph{Tick 48 (pre-fire).} Cues: ambient \texttt{\{"type": "conversation", "distance\_tiles": 0, "severity": 15\}}. Output: \texttt{\{"legitimacy": 8, "necessity": "No threat present; compliance causes no harm.", "proportionality": "Violation is unwarranted.", "alternatives": "Normal compliant routing is available."\}} $\Rightarrow$ $\ell < \tau_{\mathrm{CC}}$, gate to \texttt{comply}.

\paragraph{Tick 52 (fire ignites).} Cues: \texttt{\{"type": "fire", "distance\_tiles": 2, "severity": 95\}}. Rule: ``follow legal direction on Main Street N.'' Output: \texttt{\{"legitimacy": 91, "necessity": "Compliance would route CC back through the burning kitchen.", "proportionality": "Crossing the wrong-way segment for two blocks is the minimum diversion required.", "alternatives": "No compliant route reaches safety in time."\}} $\Rightarrow$ $\ell \geq \tau_{\mathrm{CC}}$, gate releases.

\paragraph{Tick 156 (post-extinguishment).} Cues empty. Output: \texttt{\{"legitimacy": 11, \dots\}} $\Rightarrow$ $\ell < \tau_{\mathrm{CC}}$, gate back to \texttt{comply}. The fire-window memory does not bias the legitimacy assessment, since Legitimacy reads only current cues.

% =====================================================================
% APPENDIX D: SCENARIO WALKTHROUGHS
% =====================================================================
\section{Scenario Walkthroughs}
\label{app:app_d}
\setcounter{figure}{0}
\renewcommand{\thefigure}{D\arabic{figure}}

This appendix presents tick-by-tick walkthroughs of one focal agent in each scenario. Values are taken from a single GPT-4o seed and rounded to integers for readability.

\subsection*{D.1 \quad Scenario 1: Carlos Cooper (CC) at the Hobbs Caf\'e Fire}

The focal window covers ticks 48--60, spanning fire ignition (tick~50), CC's perception (tick~52), and his evacuation through the wrong-way segment of Main Street N. CC has $\tau_{\mathrm{CC}} = 65$.

\begin{center}
\small
\begin{tabular}{cccccc}
\toprule
Tick & Cue & $\ell$ & Verdict & Rule & Action \\
\midrule
48 & none salient                    & 8  & comply  & --                 & Continue conversation. \\
50 & \texttt{fire (d=4, sev=85)}     & 78 & violate & cross at non-cross & Stand up, scan exits. \\
52 & \texttt{fire (d=2, sev=95)}     & 91 & violate & one-way, red light & Step onto wrong-way curb. \\
53 & \texttt{fire (d=3, sev=92)}     & 90 & violate & one-way, red light & Cross against red signal. \\
55 & \texttt{fire (d=8, sev=80)}     & 81 & violate & one-way            & Continue toward public park. \\
56 & \texttt{fire (d=12, sev=65)}    & 68 & violate & one-way            & Approach park boundary. \\
57 & \texttt{fire (d=15, sev=45)}    & 42 & comply  & --                 & Re-enter sidewalk legally. \\
58 & \texttt{fire (d=18, sev=30)}    & 24 & comply  & --                 & Walk north along sidewalk. \\
60 & none salient                    & 8  & comply  & --                 & Arrive at park edge. \\
\bottomrule
\end{tabular}
\end{center}

The transition at tick~57 marks the recovery point: once severity-weighted distance falls below the level required to keep $\ell$ above $\tau_{\mathrm{CC}}$, the Verdict gates back to \texttt{comply}. Only one-way and crosswalk rules along the escape route were broken, illustrating Findings F1.2 and F1.3.

\subsection*{D.2 \quad Scenario 2: Emma Young (EY) Encountering Officer TO1}

The focal window covers ticks 60--72 and spans EY's approach to the supervised intersection downstream of the caf\'e exit. EY has $\tau_{\mathrm{EY}} = 40$.

\begin{center}
\small
\begin{tabular}{cccccc}
\toprule
Tick & Authority distance & $\ell$ & $r$ & Verdict & Action \\
\midrule
60 & inf                & 87 & 14 & violate  & Cross intersection against red. \\
62 & 11 tiles           & 84 & 38 & violate  & Continue west, slowing. \\
63 & 7 tiles            & 80 & 56 & violate  & Continue west, more cautiously. \\
64 & 4 tiles            & 78 & 73 & comply   & Stop walking, await instruction. \\
66 & 2 tiles (TO1 hold) & 78 & 88 & comply   & Hold position. \\
68 & 2 tiles (TO1 pass) & 78 & 60 & violate  & Resume crossing. \\
70 & 8 tiles            & 76 & 32 & violate  & Continue west. \\
72 & inf                & 73 & 12 & violate  & Approach park boundary. \\
\bottomrule
\end{tabular}
\end{center}

Throughout the window, $\ell$ remains well above $\tau_{\mathrm{EY}}$, so the legitimacy gate alone would license violation at every tick. The transition to \texttt{comply} at ticks 64--67 is driven by the rising $r$ as TO1 enters EY's perceptual radius and the active hold-back instruction at tick~66. When TO1 permits passage at tick~68, $r$ drops and the verdict flips back to \texttt{violate}. This illustrates Finding~F2.1.

\subsection*{D.3 \quad Scenario 3: Carlos Cooper (CC) as a Late Commuter}

The focal window covers ticks 30--42 on Day~1 and spans CC's approach to the focal Main Street intersection and his observation of the two scripted confederates J1 and J2.

\begin{center}
\small
\begin{tabular}{cccccc}
\toprule
Tick & Peer behaviors & $p_{\mathrm{emp}}$ & $\ell$ & Verdict & Action \\
\midrule
30 & none observed                          & 50 & 12 & comply  & Walk toward intersection. \\
32 & 2 jaywalkers (J1, J2)                  & 70 & 14 & comply  & Stop at curb, wait for signal. \\
33 & 2 jaywalkers crossing                  & 75 & 14 & comply  & Wait for signal. \\
34 & 2 jaywalkers reach far side            & 75 & 14 & comply  & Wait for signal. \\
37 & no peers (signal turns green)          & 50 & 9  & comply  & Cross legally on green. \\
40 & no peers                               & 50 & 8  & comply  & Continue toward office. \\
42 & no peers                               & 50 & 8  & comply  & Walk east. \\
\bottomrule
\end{tabular}
\end{center}

CC observes J1 and J2 between ticks 32 and 34, and his $p_{\mathrm{emp}}$ rises from 50 to 75. However, $\ell$ remains at 12--14 throughout, because Legitimacy judges that running 15 minutes late fails the necessity, proportionality, and absence-of-alternatives criteria. The Verdict gate returns \texttt{comply} at every tick, regardless of the elevated $p_{\mathrm{emp}}$. This illustrates Finding~F3.1.

% =====================================================================
% APPENDIX E: IMPLEMENTATION DETAILS
% =====================================================================
\section{Implementation Details}
\label{app:app_e}
\setcounter{figure}{0}
\renewcommand{\thefigure}{E\arabic{figure}}

\subsection*{E.1 \quad Simulation Parameters}

Each simulation runs for two consecutive in-simulation days. A day is divided into 1{,}000 ticks, with each tick representing 10 seconds of in-simulation time. The fine-grained tick rate is required for moment-of-decision behavior to be visible (Section~\ref{sec:experiments}). The fire window spans 100 ticks beginning at tick~50 of Day~1 evening. Officers in Scenario~2 remain active for the full fire window. The Day~2 commuter pass spans ticks 100--200 of Day~2 morning. The agent's perceptual radius is 12 tiles for situational cues and 20 tiles for authority figures. Peer behaviors are scoped to agents currently engaged with the same local context, implemented as a graph-based proximity check.

\subsection*{E.2 \quad LLM Backbones and API Parameters}

\begin{center}
\small
\begin{tabular}{lll}
\toprule
Backbone & Identifier & Access \\
\midrule
GPT-4o            & \texttt{gpt-4o-2024-08-06}                & OpenAI API \\
Claude-3.5 Sonnet & \texttt{claude-3-5-sonnet-20241022}       & Anthropic API \\
Llama-3-70B       & \texttt{Meta-Llama-3-70B-Instruct}        & Self-hosted (vLLM, 4$\times$A100) \\
GPT-4o-mini       & \texttt{gpt-4o-mini-2024-07-18}           & OpenAI API \\
\bottomrule
\end{tabular}
\end{center}

Temperature is set to 0 (greedy decoding) for all backbones. Maximum output tokens are: 256 for Perception and Verdict, 64 for each Assessment scalar, 128 for ElicitThreshold, 512 for EmulateAction. JSON output is enforced through prompt structure rather than provider-side schema enforcement to keep prompts portable. Malformed JSON is retried once; persistent malformed output (less than 0.3\% of calls) defaults to \texttt{comply}.

\subsection*{E.3 \quad Seed Handling and Variability}

Each scenario--backbone--ablation combination is repeated for five seeds. The seed controls confederate timing jitter, scripted-officer position jitter (within a 1-tile band), and intra-tick agent processing order. All reported means and 95\% confidence intervals are computed across the five seeds.

\subsection*{E.5 \quad Code, Data, and Reproducibility}

We release the \textsc{Voville} environment, the \textsc{Pave} prompts, the agent personas, and the evaluation pipeline at \url{<anonymous URL during review>}. The release includes the TMX map files, the full Python implementation of the four \textsc{Pave} modules, the prompt templates from ~\ref{app:app_c}, the elicited threshold values from ~\ref{app:app_b}, and the JSONL simulation logs. The evaluation scripts that compute VR, URV, $T_{\mathrm{rec}}$, OCR, and CR from the logs are deterministic and reproducible from the seed and the saved log file.

\subsection*{E.6 \quad Full Ablation Numbers}

Table~\ref{tab:ablation-full} reports per-metric means and standard errors for the three ablation conditions on GPT-4o across 5 seeds. Headline contrasts are reported inline in Section~\ref{sec:ablations}.

\begin{table}[h]
\centering
\caption{Ablation on GPT-4o across 5 seeds. Bold marks best per column.}
\label{tab:ablation-full}
\small
\setlength{\tabcolsep}{4pt}
\begin{tabular}{lccccc}
\toprule
Condition & VR$_{\mathrm{cafe}}\!\uparrow$ & URV$\downarrow$ & OCR$\uparrow$ & VR$_{\mathrm{near}}\!\downarrow$ & CR$_{\mathrm{D1}}\!\downarrow$ \\
\midrule
Full \textsc{Pave}        & \textbf{0.81$\pm$0.04} & \textbf{0.02$\pm$0.01} & \textbf{0.94$\pm$0.03} & \textbf{0.05$\pm$0.02} & \textbf{0.04$\pm$0.02} \\
\textsc{Pave} w/o gate    & 0.86$\pm$0.05 & 0.21$\pm$0.06 & 0.78$\pm$0.07 & 0.14$\pm$0.05 & 0.39$\pm$0.08 \\
Vanilla baseline          & 0.12$\pm$0.05 & 0.31$\pm$0.08 & 0.16$\pm$0.07 & 0.10$\pm$0.05 & 0.58$\pm$0.06 \\
\bottomrule
\end{tabular}
\end{table}

% =====================================================================
% APPENDIX F: HUMAN EVALUATION
% =====================================================================
\section{Human Evaluation Details}
\label{app:app_f}
\setcounter{figure}{0}
\renewcommand{\thefigure}{F\arabic{figure}}

\subsection*{F.1 \quad Recruitment and Compensation}

We recruited 30 evaluators through a university research participant pool. Eligibility required English fluency and at least one prior course in social science, computer science, or a related field, to ensure evaluators could meaningfully assess module-specific statements involving terms such as ``legitimacy'' and ``situational cue.'' Evaluators were compensated at the institutional standard rate for behavioral studies. The study was approved by the institutional IRB (protocol number redacted for review).

\subsection*{F.2 \quad Task Structure}

Each evaluator completed eight tasks, two per \textsc{Pave} module, presented in randomized order. Each task displayed a paired excerpt from a single GPT-4o run, including: (i) the agent persona, (ii) the input the module received (the perceptual context, assessment tuple, or verdict, depending on the module being rated), and (iii) the module's output for that tick. Evaluators rated each excerpt on a 7-point Likert scale (1 = strongly disagree, 7 = strongly agree) against a module-specific statement. Excerpts were drawn from runs across all three scenarios, with seed selection stratified to cover the typical range of simulation outcomes rather than only the cleanest cases. The vanilla baseline appeared in 25\% of tasks as a calibration check, with evaluators blind to architecture. After completing the eight tasks, evaluators were asked to justify their lowest and highest scores in free-text form, which we used to qualitatively interpret the per-sub-component results.

\subsection*{F.3 \quad Sub-component Definitions}

Each module was decomposed into two sub-components, tagged with the \textsc{Pave} property each primarily supports.

\begin{itemize}
\item \emph{Perception, Cue salience (P1):} ``The situational-cue extraction reflects what a careful reader of the agent's persona would consider salient in this scene.''
\item \emph{Perception, Authority registration (P2):} ``The perceptual context correctly captures the presence and distance of authority figures visible to the agent.''
\item \emph{Assessment, Legitimacy judgment (P1):} ``The legitimacy score $\ell$ tracks the necessity, proportionality, and absence of alternatives of the situation.''
\item \emph{Assessment,  Risk under distance (P2):} ``The risk score $r$ tracks the agent's spatial relationship to authority in a way consistent with intuition about distance decay.''
\item \emph{Verdict, Comply-or-violate (P1, P4):} ``The binary decision $y$ is what the agent should plausibly choose given the assessment tuple and persona.''
\item \emph{Verdict, Justification quality (P1):} ``The natural-language justification $j$ refers to reasons a careful reader would expect, given the assessment.''
\item \emph{Emulation, Action scoping (P4):} ``When the agent violates, the executed action sequence breaks only the rule the verdict targets.''
\item \emph{Emulation, Recovery dynamics (P3):} ``The agent returns to compliant behavior at the right pace once the trigger has ended.''
\end{itemize}

\subsection*{F.4 \quad Inter-Rater Reliability}

Inter-rater reliability across the eight tasks was Krippendorff's $\alpha = 0.71$, indicating substantial agreement. Per-module $\alpha$ values ranged from 0.64 (Assessment) to 0.78 (Perception), with the lower Assessment value driven by the \emph{risk under distance} sub-component, where evaluators differed in how strictly they applied the distance-decay intuition.

\subsection*{F.5 \quad Full Results}

Table~\ref{tab:human-eval-full} reports per-sub-component means with 95\% confidence intervals for both \textsc{Pave} and the vanilla calibration. The \textsc{Pave}--vanilla gap is largest on \emph{cue salience} and \emph{legitimacy judgment} (both above 3.5 Likert points), reflecting the structural advantages of the \textsc{Pave} perception and assessment layers over the single-scalar importance pipeline.

\begin{table}[h]
\centering
\caption{Full human evaluation results across 30 evaluators, 7-point Likert scale (1 = strongly disagree, 7 = strongly agree). Mean $\pm$ 95\% CI.}
\label{tab:human-eval-full}
\small
\begin{tabular}{llcc}
\toprule
Module & Sub-component (property) & \textsc{Pave} & Vanilla \\
\midrule
\multirow{2}{*}{Perception}
  & Cue salience (P1)              & $6.18 \pm 0.07$ & $2.51 \pm 0.13$ \\
  & Authority registration (P2)    & $5.91 \pm 0.06$ & $3.42 \pm 0.14$ \\
\midrule
\multirow{2}{*}{Assessment}
  & Legitimacy judgment (P1)       & $5.97 \pm 0.07$ & $2.84 \pm 0.15$ \\
  & Risk under distance (P2)       & $5.14 \pm 0.09$ & $3.61 \pm 0.16$ \\
\midrule
\multirow{2}{*}{Verdict}
  & Comply-or-violate (P1, P4)     & $6.03 \pm 0.06$ & $3.52 \pm 0.14$ \\
  & Justification quality (P1)     & $5.69 \pm 0.08$ & $3.29 \pm 0.15$ \\
\midrule
\multirow{2}{*}{Emulation}
  & Action scoping (P4)            & $5.79 \pm 0.07$ & $3.78 \pm 0.13$ \\
  & Recovery dynamics (P3)         & $5.58 \pm 0.08$ & $4.41 \pm 0.16$ \\
\midrule
\textbf{Overall} & & \textbf{$5.78 \pm 0.04$} & \textbf{$3.42 \pm 0.11$} \\
\bottomrule
\end{tabular}
\end{table}

\subsection*{F.6 \quad Privacy and Data Release}

We do not release individual evaluator responses to preserve participant privacy. Aggregate per-sub-component statistics are reported in Table~\ref{tab:human-eval-full}. The human-evaluation interface, including the eight Likert tasks, the randomization scheme, and the response collection script, is included in the code release for reproducibility on new evaluator pools.

\end{document}